\documentclass[11pt]{article}

\newcommand\authormark[1]{\textsuperscript#1}

\usepackage{mathrsfs,amsmath}
\usepackage{chemformula}
\usepackage[mathscr]{euscript}
\usepackage{relsize}
\usepackage{bbm}
\usepackage{siunitx}
\usepackage{graphicx}
\usepackage{caption}
\usepackage{subcaption}
\usepackage[section]{placeins}
\usepackage{fancyhdr}
\usepackage{lastpage}
\usepackage[margin=1in]{geometry}

\newcommand{\unit}[1]{\ \si{#1}}

\usepackage{setspace}
\begin{document}

\title{Nanoantenna Design for Enhanced Carrier-Envelope-Phase Sensitivity}

\maketitle

\author{\noindent Drew Buckley\authormark{{1,2}}, }
\author{Yujia Yang\authormark{1}, }
\author{Yugu Yang-Keathley\authormark{2}, }
\author{Karl K. Berggren\authormark{1}, and }
\author{Phillip D. Keathley\authormark{{1,*}}}
\\
\\
\noindent \authormark{1} Research Laboratory of Electronics, Massachusetts Institute of Technology, Cambridge, MA 02139, USA. \\
\authormark{2} Department of Electrical and Computer Engineering, Wentworth Institute of Technology, Boston, MA 02115, USA\\
\authormark{*}pdkeat2@mit.edu %% email address is required



\begin{abstract}
Optical-field emission from nanostructured solids such as subwavelength nanoantennas can be leveraged to create sub-femtosecond, PHz-scale electronics for optical-field detection.  One application that is of particular interest is the detection of an incident optical pulse's carrier-envelope phase.  Such carrier-envelope-phase detection requires few-cycle, broadband optical excitation where the resonant properties of the nanoantenna can strongly alter the response of the near field in time.  Little quantitative investigation has been performed to understand how the geometry and resonant properties of the antennae should be tuned to enhance the carrier-envelope phase sensitivity and signal to noise ratio.  Here we examine how the geometry and resonance frequency of planar plasmonic nanoantennas can be engineered for enhancing the emitted carrier-envelope-phase-sensitive photocurrent when driven by a few-cycle optical pulse.  We find that with the simple addition of curved sidewalls leading to the apex, and proper tuning of the resonance wavelength, the net CEP-sensitive current per nanoantenna can be improved by $5$-$10\times$, and the signal-to-noise-ratio by $50$-$100\times$ relative to simple triangular antennas operated on resonance.  Our findings  will inform the next generation of nanoantenna designs for emerging applications in ultrafast photoelectron metrology and petahertz electronics.
\end{abstract}

%%%%%%%%%%%%%%%%%%%%%%%%%%  body  %%%%%%%%%%%%%%%%%%%%%%%%%%
\section*{Introduction}

Optical-field photoemission occurs when an optical wave’s electric field at the surface of an emitter is strong enough to drive electron tunneling from the material. Unlike perturbative photoemission mechanisms that depend on the cycle-averaged intensity of an optical pulse, optical-field emission exhibits sub-optical-cycle control of electron emission~\cite{herink_field-driven_2012, yalunin_strong-field_2011, kim_manipulation_2013, yudin_nonadiabatic_2001}, and is sensitive to changes in the incident pulse's carrier-envelope phase (CEP)~\cite{kruger2011attosecond}. Using plasmonic nanoantennas, the incident optical fields can be enhanced across a broad bandwidth near the structure’s plasmonic resonance, allowing for optical-field photoemission to be achieved using much lower incident optical pulse energies.  Recent studies have demonstrated that by driving such nanoantennas with few-cycle optical pulses, an appreciable CEP-sensitive photocurrent can be generated using only picojoules of energy at repetition rates approaching \SI{100}{\mega\hertz}~\cite{rybka_sub-cycle_2016, putnam_optical-field-controlled_2017, keathley_phillip_d._antiresonant-like_2019, ludwig2019sub}.  With proper engineering, such nanoantenna-enhanced photodetectors could lead to compact and integratable CEP-sensitive detectors that provide shot-to-shot CEP stabilization and control for compact frequency combs and few-cycle optical sources~\cite{yang2019light}.  

Prior works investigating CEP-sensitive photoemission from plasmonic antennas have only explored simplified, triangular geometries, and have yet to investigate how the nanoantenna resonance frequency impacts the CEP-sensitive photocurrent.  In this work, we investigated the impact of nanoantenna shape and resonance frequency on the CEP-sensitive photocurrent generated in the optical-field emission regime when the antennas are illuminated by a fixed source. By introducing curved sidewalls leading to the antenna apex, we found that the enhanced intensity at the tip apex can be increased by roughly $2\times$ compared to triangular nanoantennas while maintaining the same enhanced pulse duration.  Following this, we systematically investigated the impact of the antenna resonance frequency on the CEP-sensitive photocurrent per antenna, finding that the maximum CEP-sensitive photocurrent is, in general, achieved when the resonance of nanoantenna is roughly aligned with the long-wavelength edge of the source spectrum.  This condition strikes a balance between field enhancement and pulse duration as measured in the number of optical cycles of the enhanced electric field waveform at the antenna apex.  

While it was noted in prior work~\cite{putnam_optical-field-controlled_2017, yang2019light} that tuning the nanoantenna resonance away from the illuminating optical pulse's central wavelength results in shorter enhanced pulse durations, and thus improved CEP-sensitivity, only detuning toward shorter wavelengths was explored.  Our findings show that significant improvements in the overall CEP-sensitive signal and signal to noise ratio could be achieved by detuning towards longer wavelengths.  With a combination of the geometric and resonant tuning we explore here, the net CEP-sensitive current per nanoantenna can be improved by $5$-$10\times$, and the signal-to-noise-ratio by $50\times$ to greater than $100\times$ relative to simple triangular antennas operated on-resonance depending on the peak incident electric field strength.  

\section*{Simulation Overview}

\begin{figure}[h]
		\centering
		\includegraphics[width=\textwidth]{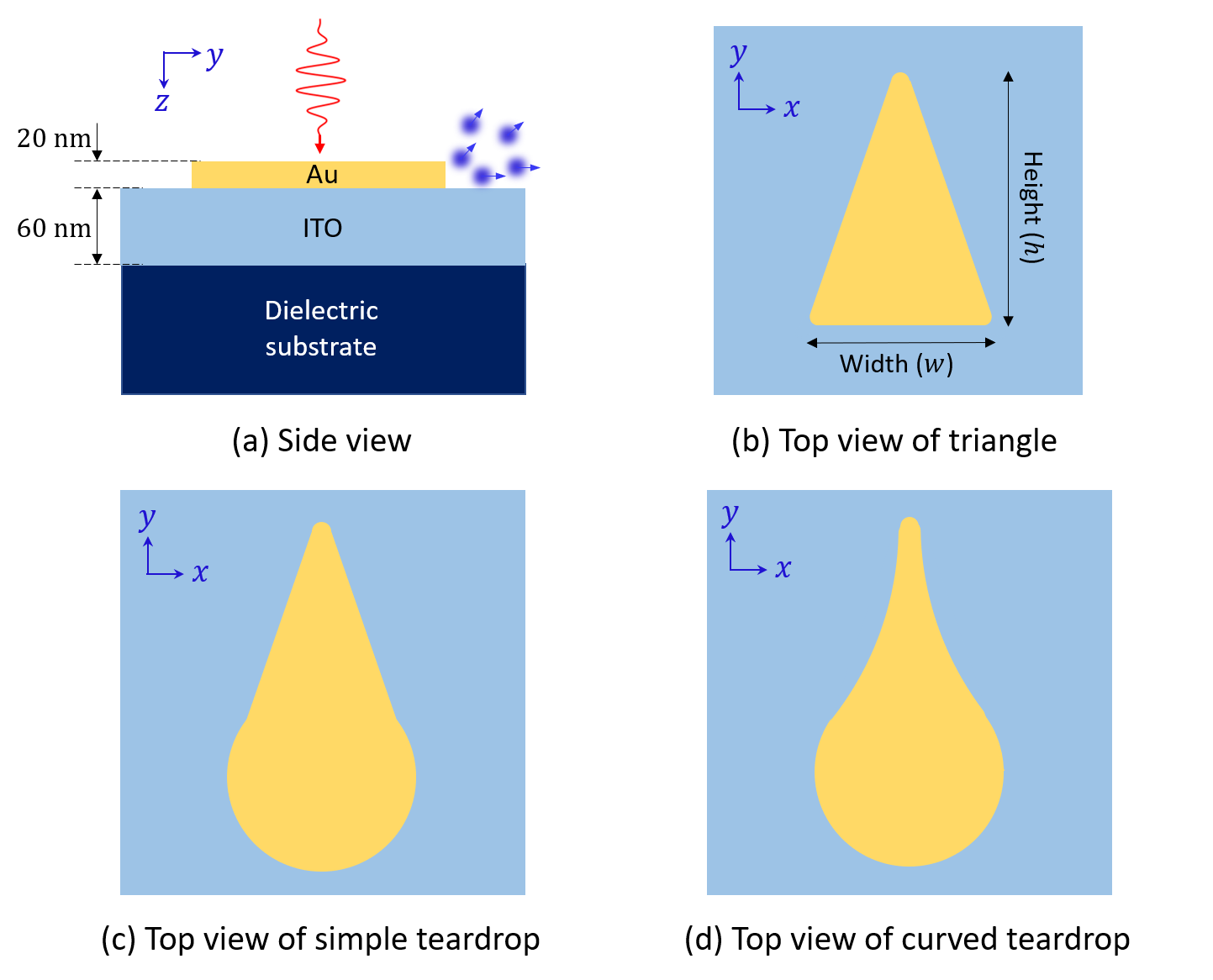}
		\caption{Illustrations of the simulation cell and simulated nanotip geometries. (a) A side view of the simulation cell showing incident waveform, nanoantenna, conductive ITO layer, and dielectric substrate (assumed to be sapphire throughout this work).  The thicknesses of the nanoantennas (\SI{20}{\nano\meter}) and the ITO layer (\SI{60}{\nano\meter}) were held fixed for all simulations.  (b) Top view of the triangle geometry. (c) Top view of the simple teardrop geometry. For this shape, the top portion of the antenna leading to the apex was identical to the triangle geometry, but the base has been changed to be a section of a cylinder.  (d) Top view of the curved teardrop geometry.  For this geometry, curvature was added to the sidewalls leading to the nanoantenna apex (see text and Appendix A).  For each case, the tip apex radius of curvature was held fixed at \SI{10}{\nano\meter}.}
		\label{fig:simulation_cell}
\end{figure}

An example simulation cell is shown in Fig.~\ref{fig:simulation_cell}.  A gold nanoantenna with a thickness of \SI{20}{\nano\meter} sits on a 60-nm thick layer of indium tin oxide (ITO), which is often used in experiments to prevent substrate charging.  Underneath the nanoantenna and ITO layer was a semi-infinite sapphire substrate. These materials and thicknesses were held fixed in this study as our focus was on the impact of the antenna shape and resonance frequency.  The incident optical pulses were modeled as plane wave pulses propagating orthogonal to the surface of the nanoantenna, with a polarization axis aligned as indicated in Fig.~\ref{fig:simulation_cell}-a.  

We evaluated three variations of nanoantenna geometry with the same fixed radius of curvature of \SI{10}{\nano\meter} at the apex as illustrated in Figure~\ref{fig:simulation_cell}: (1) a triangle geometry (seen in Figure~\ref{fig:simulation_cell}-b) that has a height-to-width ratio ($h/w$) equal to $4/3$; (2) a simple teardrop geometry (seen in Figure~\ref{fig:simulation_cell}-c) that was formed by attaching a cylindrical base to the bottom of the triangular geometry such that the diameter of the cylinder of length $w$ is aligned with the base of the triangle; and (3) a curved teardrop geometry (seen in Figure~\ref{fig:simulation_cell}-d) that was formed by applying curvature to the simple teardrop geometry along each side leading to the antenna's apex.  The aspect ratio of the triangle antenna, which was the starting point for each design, was held fixed throughout this work, but of course could be systematically studied as a parameter of interest in future investigations.   We note, however, that it is important that the devices are asymmetric to maintain CEP-sensitivity, as it was shown in Ref.~\cite{putnam_optical-field-controlled_2017} that symmetric antennas result in a doubly-rectified current response with respect to the incident optical field, thus resulting in negligible CEP-sensitive photocurrent.  Furthermore, in Refs.~\cite{pettine_continuous_2020, hobbs_mapping_2017} it was observed that hot electron from the body of symmetric rod-like nanoantennas and nanospheres can contribute significantly to the emitted photocurrent, unlike asymmetric devices such as triangular nanoantennas where optical-field-driven photocurrent from the emitter apices dominated photoemission.  Such hot-electron emission is CEP-insensitive and is thus detrimental to CEP-sensitivity and signal to noise ratio of the antennas.  

\FloatBarrier

We used the following computational procedure to study the photoemitted charge from a nanoantenna in the optical-field emission regime. First we used finite-difference time-domain electromagnetic simulation~\cite{oskooi_meep_2010} to determine the transfer function over a bandwidth ranging from $\SI{0.8}{\micro\meter}$ to $\SI{3.0}{\micro\meter}$ at each point on the structure’s surface. Using these transfer functions, we calculated the surface field when illuminating the nanoantenna with an ultrafast optical waveform. The focus of this work was to determine how geometric properties of the nanoantenna can be altered to improve device performance assuming a fixed optical source, which was taken to be a $\cos^2$-pulse having a central wavelength of $\SI{1.177}{\micro\meter}$ and a full-width at half-maximum (FWHM) of $\SI{10}{\femto\second}$ for all of the simulations in this manuscript. These source parameters were chosen to match those of recent experiments~\cite{putnam_few-cycle_2019, putnam_optical-field-controlled_2017, keathley2019vanishing, yang2019light} making it easier to benchmark the validity of our modeling results.  However, we note that the conclusions of this work are general in nature, and should apply equally well to other source wavelengths.  

After determining the temporal fields across the surface of the nanoantenna, we modeled the optical-field-driven tunneling by applying a quasi-static tunneling model whereby the instantaneous electron emission rates were calculated using a physically complete Fowler-Nordheim-type equation~\cite{fowler_electron_1928} following the treatment as discussed in~\cite{forbes_description_2008}.  To ensure the validity of this quasi-static tunneling approximation, we only considered field strengths such that the peak enhanced field at the antenna apex was sufficiently high to ensure that the Keldysh parameter $\gamma < 1$~\cite{keldysh1965ionization}~\cite{yalunin_strong-field_2011}.  The emitted current density, which was calculated in units of \si{\ampere\per\nano\meter\tothe{2}} at each location along the nanoantenna surface, was taken to be
\begin{equation} \label{eq:fn}
	J(F) = a\phi^{-1} \Phi(-F)F^2 \exp\left[ \frac{-v b \phi^{3/2}}{F} \right] \mbox{,}
\end{equation}
where the variable $v$ is the adjustment factor, $F$ the electric field strength normal to the nanoantenna surface in \si{\volt\per\nano\meter}, $\Phi$ is the Heaviside function, and $\phi$ the work function in \si{\eV}. The adjustment factor was defined by
\begin{equation} \label{eq:fn_adj}
	v \approx 1 - f + \frac{1}{6}f\ln f \mbox{,}
\end{equation}
where $f$ is the barrier field which was approximated to 
\begin{equation} \label{eq:fn_lil_f}
	f \approx \left(1.44 \frac{\unit{eV^2 \ nm}}{V} \right) \left( \frac{F}{\phi^2} \right) \mbox{.}
\end{equation}
The work function of gold was taken to be $\SI{5.1}{\eV}$ for all of the simulations in this work.  The scaling factors $a$ and $b$ were taken to be $1.5 \times 10^{-6}$~\si{\ampere\electronvolt\per\volt\tothe{2}} and $\SI{6.8}{\volt\per\electronvolt\tothe{3/2}\per\nano\meter}$, respectively.  The instantaneous emitted photocurrent was then calculated by integrating the current density $J$ over the entire surface of the nanoantenna exposed to air, ignoring the surface adjacent to the underlying ITO.  

\begin{figure}[h]
		\centering
		\includegraphics[width=\textwidth]{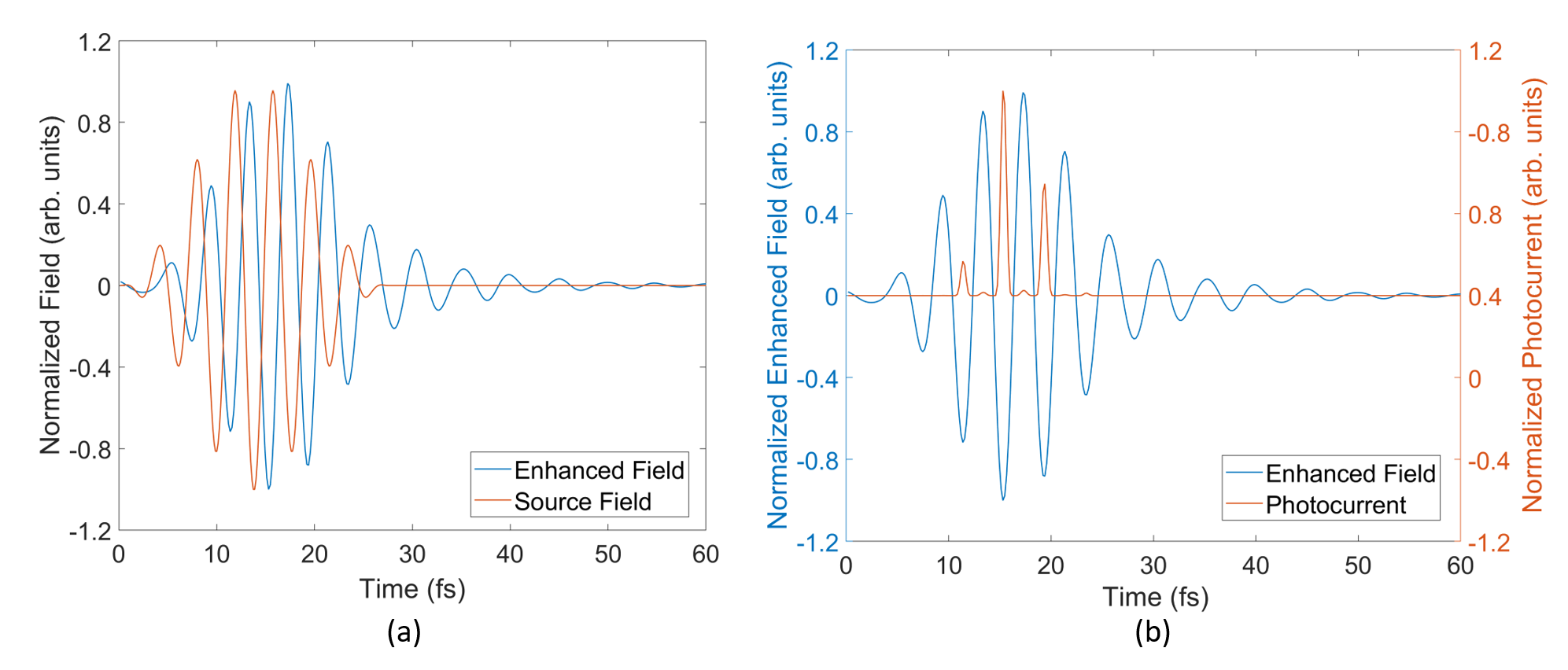}
		\caption{a) Comparison of a normalized source waveform and a normalized enhanced waveform at the apex of a triangular nanoantenna with height of \SI{240}{\nano\meter} and resonance wavelength of \SI{1.1}{\micro\meter}. Note that the peak field strength of the enhanced waveform is roughly $30\times$ that of the source waveform.  The enhanced waveform experiences both a phase shift and a change in shape due to the resonant response of the nanoantenna. b) Comparison of the enhanced waveform (blue) at the apex of the same nanoantenna as in (a) against the total instantaneous photoemission current (red). The peak incident field strength was chosen to be \SI{1}{\volt\per\nano\meter} for this calculation.  Small amounts of photocurrent can be seen for positive half-cycles; these are caused by out-of-phase electron emission from the bottom of the antenna.}
		\label{fig:fields_and_current}
\end{figure}

In Fig.~\ref{fig:fields_and_current}-a we show a representative relationship between the source waveform and the excited waveform at a nanoantenna apex.  The nanoantenna was triangular in shape with a height of \SI{240}{\nano\meter} and a resonance wavelength of $\lambda_\text{res} = \SI{1.1}{\micro\meter}$.  The peak strength of the enhanced field (blue) is $\approx 30\times$ greater than that of the source field (red) for the case shown, so both have been normalized for comparison.  Due to the resonant properties of the nanoantenna, the near-field waveform is altered in time relative to the incident waveform, with a tail following the driven response at longer times~\cite{csete_few-cycle_2020}.  As we will discuss later, with proper resonance tuning, the bandwidth of the local field at the antenna apex can be increased and the central wavelength slightly shifted thereby altering the duration of the local electric field waveform in number of optical cycles, which can significantly enhanced the CEP-sensitive photoemission per antenna.  

In Fig.~\ref{fig:fields_and_current}-b we plot the normalized enhanced waveform at the tip apex (blue) alongside the instantaneous photocurrent found by integrating the current density over the entire surface of the nanoantenna.  For the current calculation, the incident peak field strength was taken to be \SI{1}{\volt\per\nano\meter}, resulting in a peak enhanced field of roughly \SI{30}{\volt\per\nano\meter}.  Note that most of the current arises when the field at the apex is strong and negative near the central portion of the pulse as the current is dominantly emitted from the region of the apex.  This rectified, sub-cycle current response is essential to achieving high CEP-sensitivity~\cite{putnam_optical-field-controlled_2017}.  The emitted charge yield per pulse $Q$ by integrating this instantaneous photocurrent over all time.  

%An example calculation of the instantaneous photocurrent is shown in Fig.~\ref{fig:fields_and_current}-b.    

\FloatBarrier

\begin{figure}[h]
		\centering
		\includegraphics[width=0.8\textwidth]{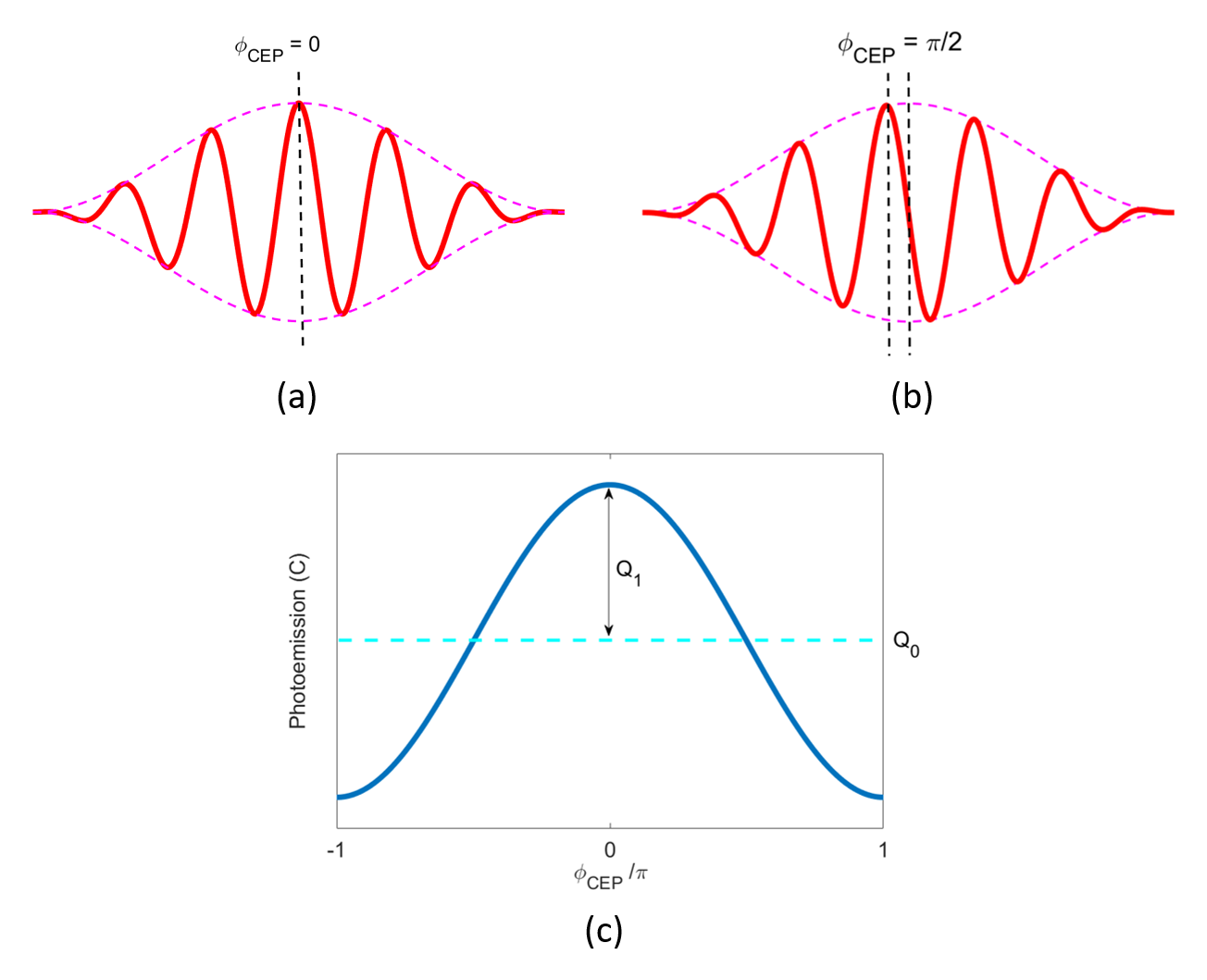}
		\caption{Definition of CEP and CEP-sensitive photocurrent.  a,b) An illustration of two optical pulses with identical center wavelength and duration, but different CEP. c) Illustration of how charge emission per pulse changes as CEP is shifted. The light blue dashed line represents average charge emitted per pulse ($Q_0$). The blue solid curve represents charge emitted per pulse as a function of CEP. The CEP dependent charge emission ($Q_1$) is defined as the amplitude of the first harmonic of the CEP-sensitive current.}
		\label{fig:cep_and_cep_sensitivity}
\end{figure}
\FloatBarrier

In the final step, we calculated $Q$ over a $2\pi$-sweep of the CEP and a range of field strengths of the incident optical waveform.  Figure~\ref{fig:cep_and_cep_sensitivity}-a,b shows how a shift in CEP changes the electric field profile of the optical waveform. Figure~\ref{fig:cep_and_cep_sensitivity}-c shows how the emitted charge varies with CEP at a fixed field strength. Throughout the remainder of this work, we will refer to two characteristic components of the emission signal as labeled in Figure~\ref{fig:cep_and_cep_sensitivity}-c: (1) the average total charge per pulse denoted as $Q_0$; and (2) the amplitude of the CEP-dependent charge per pulse $Q_1$, which indicates to the sensitivity of the emitted charge to the shift of CEP. In our calculations, we specifically define $Q_1$ as the amplitude of the first harmonic of the CEP-dependent charge signal as the emitted charge variation with CEP is well-represented by a sinusoidal function.  It is important to take into account both $Q_1$ and $Q_0$ as recent measurements have shown that the signal-to-noise ratio (SNR) of the CEP-sensitive photocurrent from such nanoantenna devices is shot-noise limited~\cite{yang2019light}.  In the shot-noise limit, the noise grows as the square-root of the total charge, and it can be shown that the power SNR of the CEP-sensitive photoemission measurement is directly proportional to $Q_1^2/Q_0$.  

\section*{Results and Discussion}

\subsection*{Emitter Shape Comparison}

In this section we explore the effect of nanoantenna's geometry on the excited electric field profile and photoemission. For simplicity, we only show results from each geometry for a fixed resonance wavelength of $\lambda_{\text{res}} = 1.1$ \si{\micro\meter}.  However, we note that other resonance wavelengths were tested, all following the same qualitative trend as observed in the case of $\lambda_{\text{res}} = 1.1$ \si{\micro\meter}.  

\begin{figure}[h]
		\centering
		\includegraphics[width=\textwidth]{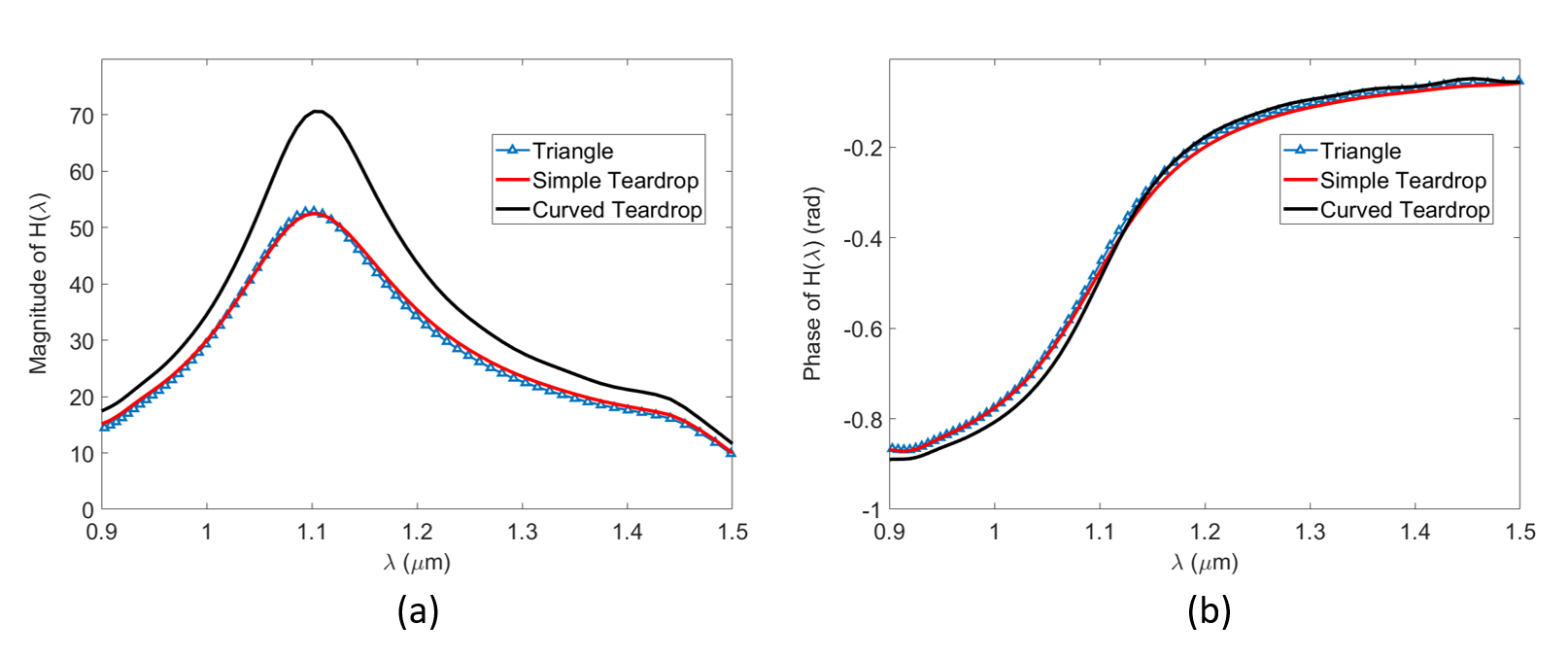}
		\caption{Electric field transfer function $H(\lambda)$ at the tip apex for all three geometries. a) Field enhancement spectrum $|H(\lambda)|$. b) The phase response spectrum $\angle(H(\lambda))$.  Note that the shape of the field enhancement and phase response spectra are almost identical for all three geometries.  This indicates that the improved field enhancement observed for the curved teardrop is due to an enhanced lightning-rod effect.  This is advantageous for ultrafast applications where braod field enhancement spectra with relatively flat phase responses are desired.}
		\label{fig:transfer_function_1p1_micron}
\end{figure}

In Figure~\ref{fig:transfer_function_1p1_micron}, we plot the electric field transfer function $H(\lambda) = E_{\text{tip}}(\lambda)/E_{\text{inc}}(\lambda)$ evaluated at the tip, where $\lambda$ is the incident optical wavelength, $E_\text{inc}$ the incident electric field of the optical waveform, and $E_{\text{tip}}$ is the excited field at the tip of the nanoantenna.  The amplitude and phase of the transfer function are plotted separately.  The magnitude of $H(\lambda)$, shown in Figure~\ref{fig:transfer_function_1p1_micron}-a, represents apex field enhancement for each sinusoidal frequency component of an incident optical waveform. The phase of $H(\lambda)$, shown in Figure~\ref{fig:transfer_function_1p1_micron}-b, represents the phase shift of each wavelength component due to the interaction with the nanoantenna. While the triangle and simple teardrop have approximately the same apex field enhancement across the spectrum, the curved teardrop exhibits a significant increase in field enhancement. For the case of the curved teardrop, the maximum field enhancement at $\lambda = \lambda_\text{res}$ is $71\times$, which is approximately $1.4 \times$ that of the triangle and simple teardrop geometry. The profiles of the phase and normalized magnitude of $H(\lambda)$  of the three geometries are almost identical, resulting in nearly identical normalized electric field waveform at the apex of the three geometries.  This indicates that the improved field enhancement of the curved teardrop arises due to a geometrically-enhanced lightning-rod effect rather than an improvement of the quality factor of the antenna rsonance.  As we will see, this is advantageous for ultrafast applications such as CEP detection where one desires very broadband field enhancements with a relatively flat phase response.    

\FloatBarrier

\begin{figure}[h]
		\centering
		\includegraphics[width=\textwidth]{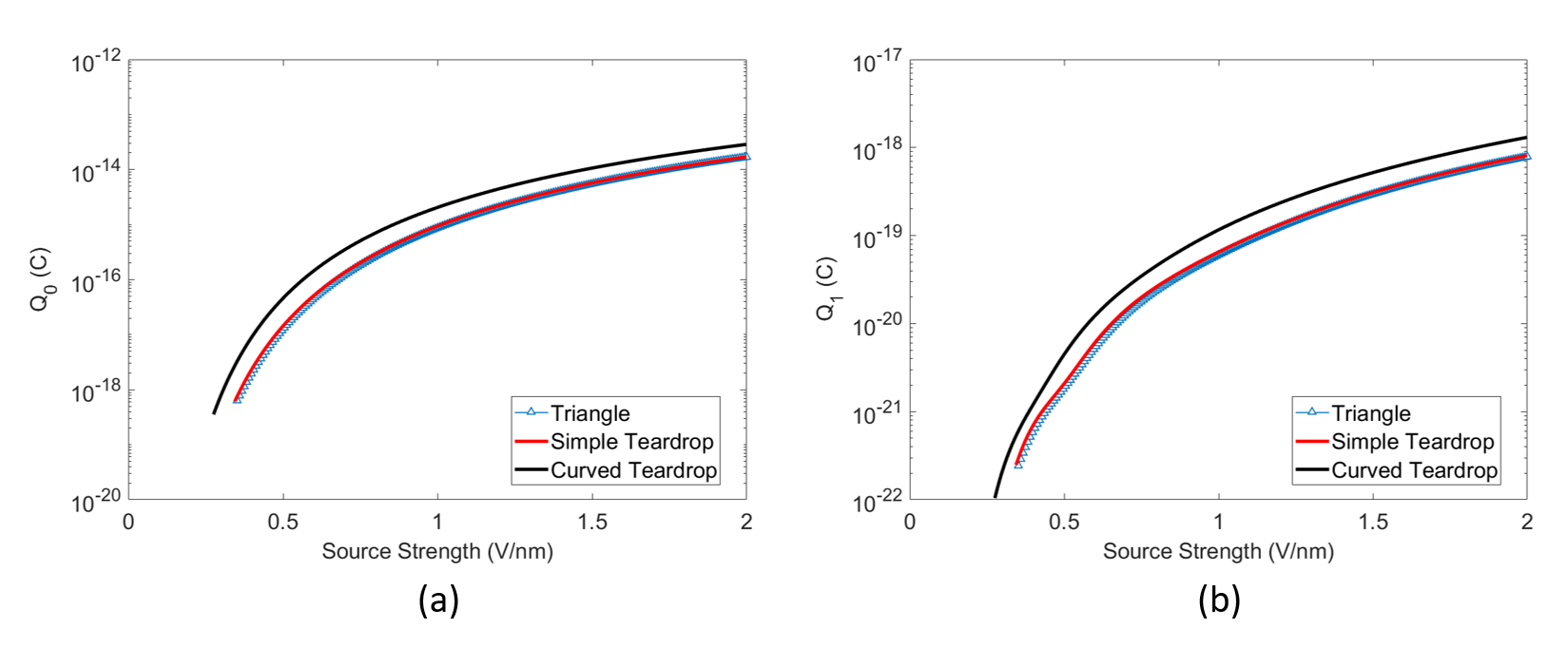}
		\caption{Charge emission results for each geometry.  For each geometry, the resonance frequency was $\lambda_{\text{res}} = \SI{1.1}{\micro\meter}$. a) Average total charge emission per pulse $Q_0$.  b) The CEP-dependent emission per pulse $Q_1$.  The average total and CEP-sensitive photocurrent responses from the triangle and simple teardrop were almost identical, while the response from the curved teardrop was up to $4.2\times$ larger over the tested range as a result of the enhanced lightning-rod effect.}
		\label{fig:charge_emission_1p1_micron}
\end{figure}
\FloatBarrier

To see the impact of the geometry on the average total and CEP-sensitive charge emission, in figure~\ref{fig:charge_emission_1p1_micron} we plot  $Q_0$ (a) and  $Q_1$ (b) over a range of peak field strengths of the incident optical waveform for each emitter shape.  The geometrically-enhanced lightning-rod effect of the curved teardrop increased the enhanced intensity at the nanoantenna apex by roughly a factor of two, resulting in an increase of both $Q_0$ and $Q_1$ ranging from $1.7\times$ to $4.2\times$ relative to the other two geometries over the range of tested field strengths.  This improved intensity enhancement is significant as it means the required incident intensity needed for the same CEP-sensitive emission per antenna is halved.  

\begin{figure}[h]
		\centering
		\includegraphics[width=\textwidth]{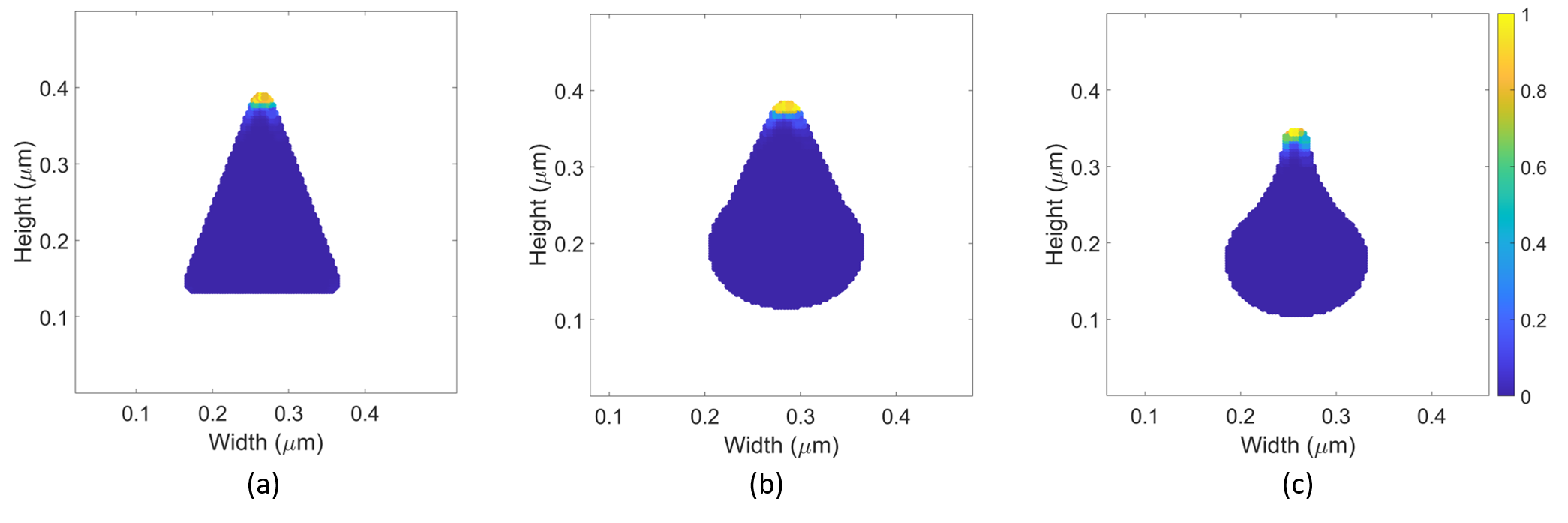}
		\caption{Pseudocolor plot of the normalized current density from each nanantenna geometry.  For these plots, the incident field strength was taken to be \SI{1}{\volt\per\nano\meter} a) Triangle geometry. b) Simple teardrop geometry. c) Curved teardrop geometry.  The tip apex contributed to more than 97.8\% of the total calculated current for the triangular geometry, 99.1\% for the simple teardrop geometry, and 99.7\% for the curved teardrop geometry.}
		\label{fig:pcolor_current_density}
\end{figure}

While the increased the CEP-sensitive and average total photocurrent response was expected for the curved teardrop due to the observed increase in field enhancement across the entire spectrum of the incident pulse, we were surprised to find that the photocurrent from the triangular and simple-teardrop geometries were almost identical.  Initially, we suspected that the simple teardrop geometry might show an improved CEP-sensitive response relative to the triangular geometry due to the removal of the two corners at the bottom of the triangle. This is because these corners lead to increased fields at the back side of the antenna relative the rounded geometry of the simple teardrop.  The fields at the back side of the antenna are $180^\circ$ out of phase with the field at the apices.  This $180^{\circ}$ phase shift in field results in a $180^\circ$ shift in the CEP-sensitive photoemission from the back side of the antenna.  Thus, a strong amount of such back-side emission would cause a suppression of $Q_1$ as discussed in Refs.~\cite{putnam_optical-field-controlled_2017, keathley2019vanishing}.  However, due to the high degree of nonlinearity of the photoemission process, we find that this out-of-phase emission from the back sides of the antennae is negligible for all three tested geometries.  To show this, we plot the normalized emitted current density from each antenna shape for an incident field strength of \SI{1}{\volt\per\nano\meter} in Figure~\ref{fig:pcolor_current_density}.   The percentage of the total current contribution coming from the tip apex was calculated to be 97.8\% for the triangular antenna, 99.1\% for the simple teardrop, and 99.7\% for the curved teardrop. Given the identical geometry for the top half of the triangle and simple teardrop, this explains why the photocurrent from the triangular and curved teardrop geometries were so similar.  It also supports findings in Ref.~\cite{hobbs_mapping_2017} where photoresist was used to map optical-field emission from triangular nanoantennas, and minimal resist exposure was obsurved from the back corners of the triangular nanoantennas.   

While an exhaustive geometrical optimization study is beyond the scope of this work, we feel that our findings motivate such an investigation in order to determine the extent to which the total average and CEP-sensitive photocurrent could be enhanced by simply altering the nanoantenna shape.

\FloatBarrier

\subsection*{Resonance Wavelength Tuning}

% -- Commented out for now -- 
% -- Remove if not desired after next complete read --
%Next, we examine how field enhancement and photoemission are affected by $\lambda_\text{res}$.  Following on the results of the previous section, we found that for every $\lambda_\text{res}$ tested, the curve teardrop always produced larger field-enhancement factors while maintaining a nearly identical normalized electric field waveform at the apex of the nanoantenna.  This means that other than a slight shift in intensity as shown in Figure~\ref{fig:charge_emission_1p1_micron}, the dependence of the photoemission on $\lambda_\text{res}$ was essentially identical for each geometry.  As such, in this section we only discuss the curved teardrop results for simplicity.   

In this section, we examine the dependence of the CEP-dependent photoemission on $\lambda_\text{res}$ and source field strength. We have restricted this study to the curved teardrop for simplicity since all three antenna geometries we observed a similar resonant response with the only effective difference being a nearly constant, multiplicative intensity enhancement due to an enhanced lightning-rod effect.

To tune the resonance wavelength $\lambda_\text{res}$, we changed the height $h$ and the width $w$ of each nanoantenna proportionally, maintaining the same aspect ratio as described earlier and in Appendix A. Figure~\ref{fig:curved_teardrop_transfer_function} shows the transfer functions $H(\lambda)$ for different $\lambda_\text{res}$ values against the normalized spectrum of the source electric field intensity. As $\lambda_\text{res}$ becomes larger, the on-resonance field enhancement becomes greater. At $\lambda_\text{res} = \SI{1.76}{\micro\meter}$, we achieve a maximum on-resonance field enhancement of $110 \times$. At $\lambda_{\mbox{res}} = \SI{1.05}{\micro\meter}$, we see a minimum on-resonance field enhancement of $65 \times$. The dependence of the on-resonance field enhancement on the resonance wavelength is also plotted in Figure~\ref{fig:curved_teardrop_transfer_function} (dashed blue line). 

\begin{figure}[h]
		\centering
		\includegraphics[width=0.9\textwidth]{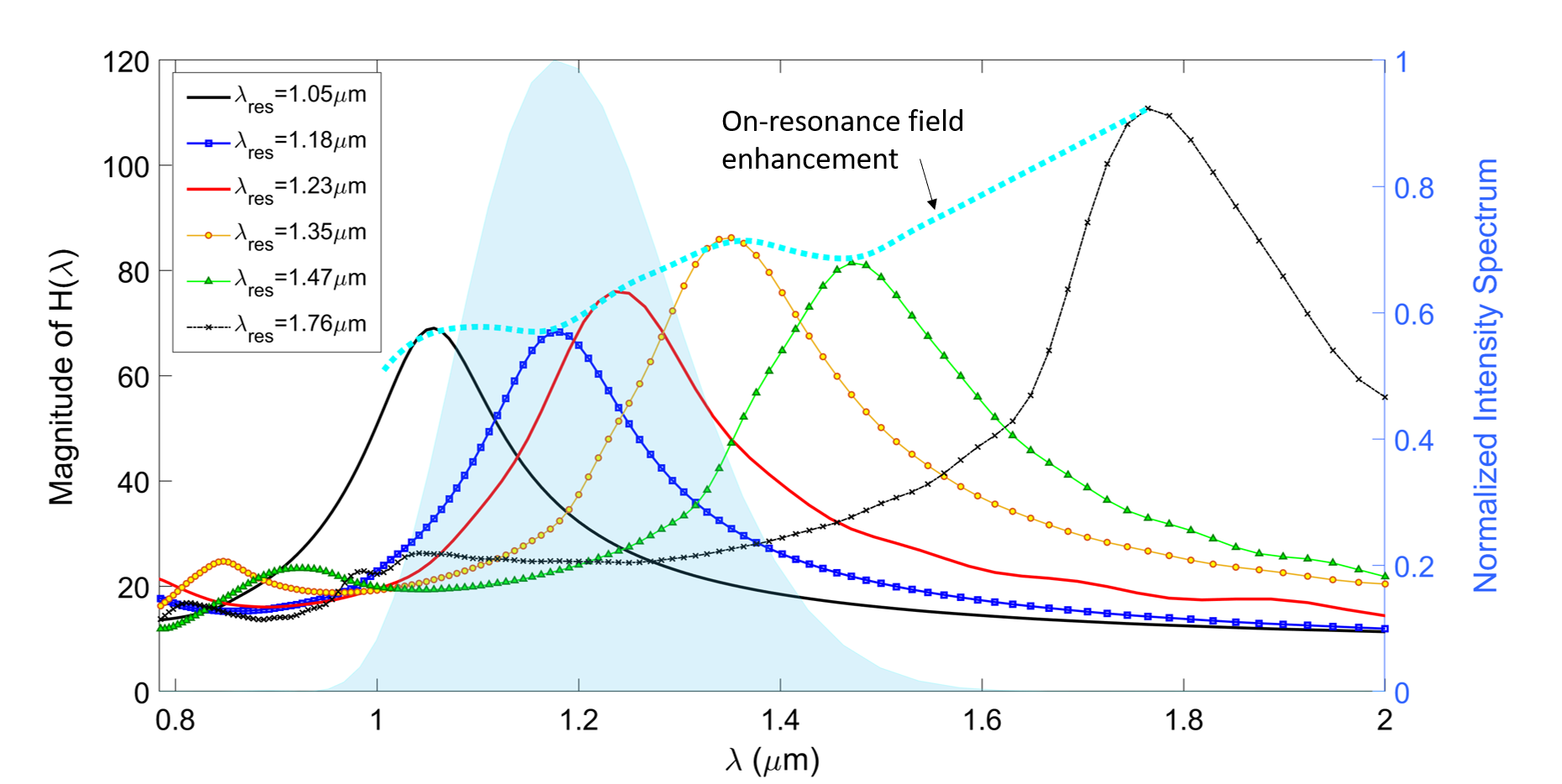}
		\caption{Magnitude of the transfer function (\textit{i.e.} the field-enhancement spectrum) of the curved teardrop as $\lambda_\text{res}$ is changed for selected values. The blue dashed line shows the field enhancement at $\lambda_\text{res}$ for all tested cases over a much finer spacing.  Aside from a few momentary dips, increasing $\lambda_\text{res}$ generally results in an increase of the on-resonance field enhancement and a larger bandwidth of the transfer function.  The shaded blue region shows the spectrum of the incident optical pulse used for the photocurrent simulations.}
		\label{fig:curved_teardrop_transfer_function}
\end{figure}
\FloatBarrier

\begin{figure}[h]
		\centering
		\includegraphics[width=0.6\textwidth]{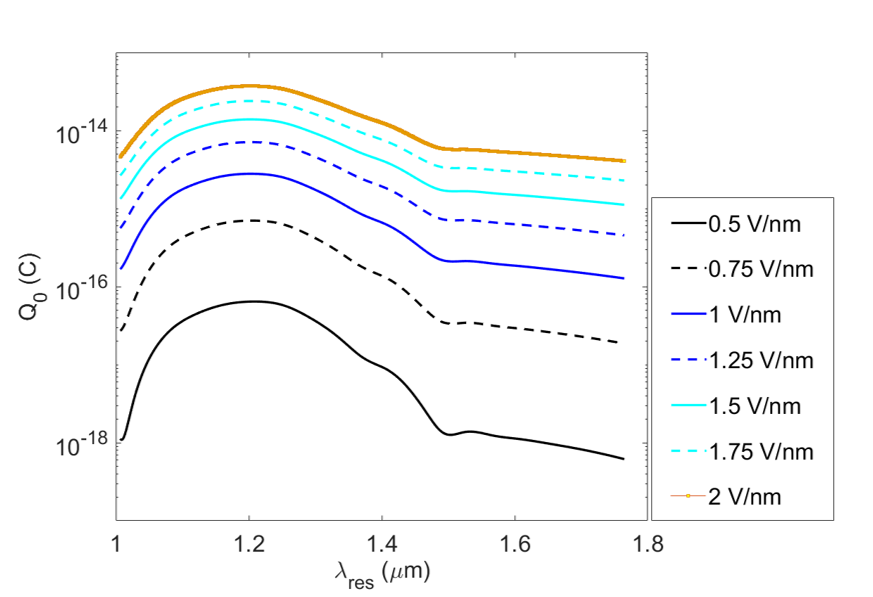}
		\caption{Average total photocurrent $Q_0$ from the curved teardrop nanotip at several source strength samples as a function of resonance wavelength.  As expected, the peak in average total photocurrent occurs when the resonance is aligned with the central wavelength of the incident optical pulse.  Furthermore, the shape of the response curve is not significantly dependent on the incident field strength.  Note, however, that the response is asymmetric with respect to $\lambda_{\text{res}}$, falling off more slowly as $\lambda_{\text{res}}$ shifts to longer wavelengths.}
		\label{fig:average_total_emission_curved_teardrop}
\end{figure}

We start by investigating the impact of the resonance wavelength on the average total charge emission $Q_0$.  Figure~\ref{fig:average_total_emission_curved_teardrop} plots $Q_0$ as a function of wavelength for various selected peak field strengths. For each field strength, the dependence of $Q_0$ on wavelength remains fairly similar.  We see that  $Q_0$ reaches the maximum at around $\approx \SI{1.2}{\micro\meter}$, near the central wavelength of the source pulse. However, as the resonance is detuned away from the central wavelength of the source pulse, the decrease in $Q_0$ is not symmetric. Resonance wavelengths shorter than $\SI{1.2}{\micro\meter}$ experience a sharp decrease in the average total charge emission as resonant wavelength decreases, while $Q_0$ decreases more gradually for resonance wavelengths longer than $\approx \SI{1.2}{\micro\meter}$. This asymmetry in the dependence on resonance wavelength can be attributed to both a narrower bandwidth of the transfer function and the decrease of on-resonance field enhancement as $\lambda_{\text{res}}$ is decreased (see Fig.~\ref{fig:curved_teardrop_transfer_function}.

\FloatBarrier

\begin{figure}[h]
		\centering
		\includegraphics[width=\textwidth]{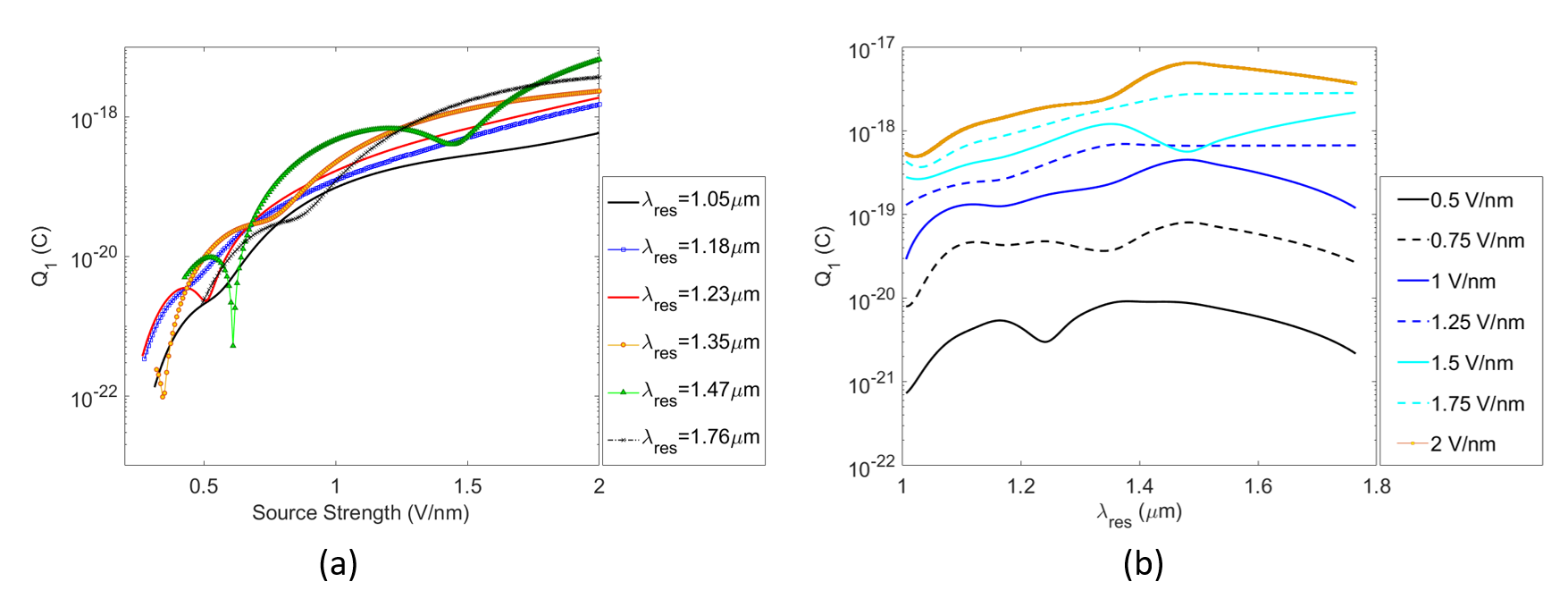}
		\caption{CEP-dependent charge emission $Q_1$. a) CEP-dependent charge emission as a function of peak source strength for selected resonance wavelengths. Vanishing points are observed for several resonance wavelength values; for instance near \SI{0.6}{\volt\per\nano\meter} for $\lambda_{\text{res}} = \SI{1.47}{\micro\meter}$ b) CEP-dependent charge emission as a function of resonance wavelength for selected peak source strength values.  The CEP-dependent photocurrent is generally larger and has a flatter response for larger values of $\lambda_{\text{res}}$.  However, due to the vanishing points, the dependence of $Q_1$ on the resonance wavelength can experience sudden changes making it difficult to pinpoint a fixed optimum resonance condition for all field strengths.}
		\label{fig:cep_dependent_charge_curved_teardrop}
\end{figure}

The dependence of the CEP-dependent charge emission $Q_1$ on both incident peak field strength and resonance wavelength is shown in Figure~\ref{fig:cep_dependent_charge_curved_teardrop}.  For certain resonance wavelengths -- most notably for longer resonance wavelengths -- sudden dips in charge emission are observed at particular peak field strengths of the incident pulse as shown in Figure~\ref{fig:cep_dependent_charge_curved_teardrop}a. These dips have been investigated both theoretically and experimentally in Ref.~\cite{keathley2019vanishing} where they were coined vanishing points. For example, a vanishing point is observed at a source strength of approximately $\SI{0.6}{\volt\per\nano\meter}$ for the $\SI{1.47}{\micro\meter}$ curved teardrop.  Ref.~\cite{keathley2019vanishing} noted that these vanishing points are highly sensitive to even minute changes in the surface-enhanced electric field waveform at the emitter tip, which explains why the source field strengths at which they occur are so sensitive to the resonance wavelength.  The presence of these vanishing points also explains why in Figure~\ref{fig:cep_dependent_charge_curved_teardrop}b, the dependence of $Q_1$ on the resonance wavelength experiences sudden changes in shape as the peak field strength is increased.

\FloatBarrier

This complex dependence of $Q_1$ on the incident field strength and resonance wavelength make it difficult to pinpoint an optimal resonance condition for maximizing CEP-sensitive photocurrent across all incident peak field strengths.  However, in general, longer resonance wavelengths tend to lead to larger net CEP-sensitive photocurrents, with the maximum value obtained near $\lambda_{\mbox{res}} \approx 1.5 \si{\micro\meter}$, which corresponds to the long-wavelength edge of the source spectrum (see Fig.~\ref{fig:curved_teardrop_transfer_function}-a).  This condition strikes a balance between the peak field enhancement obtained in the time-domain at the tip apex and the duration of the pulse in number of optical cycles.  

%-- Delte this in latest version of manuscript...
%The reason why a larger net CEP-sensitive currents $Q_1$ is obtained when the resonance wavelengths of the nanoantennas are aligned with the longer-wavelength of the source spectrum is because of a trade-off between field enhancement, which increases the average total photoenmission $Q_0$, and fewer cycles in the optical waveform, which increases CEP-sensitivity $Q_1/Q_0$ as plotted in Figure~\ref{fig:SNR_curved_teardrop}.  

\begin{figure}[h]
		\centering
		\includegraphics[width=0.6\textwidth]{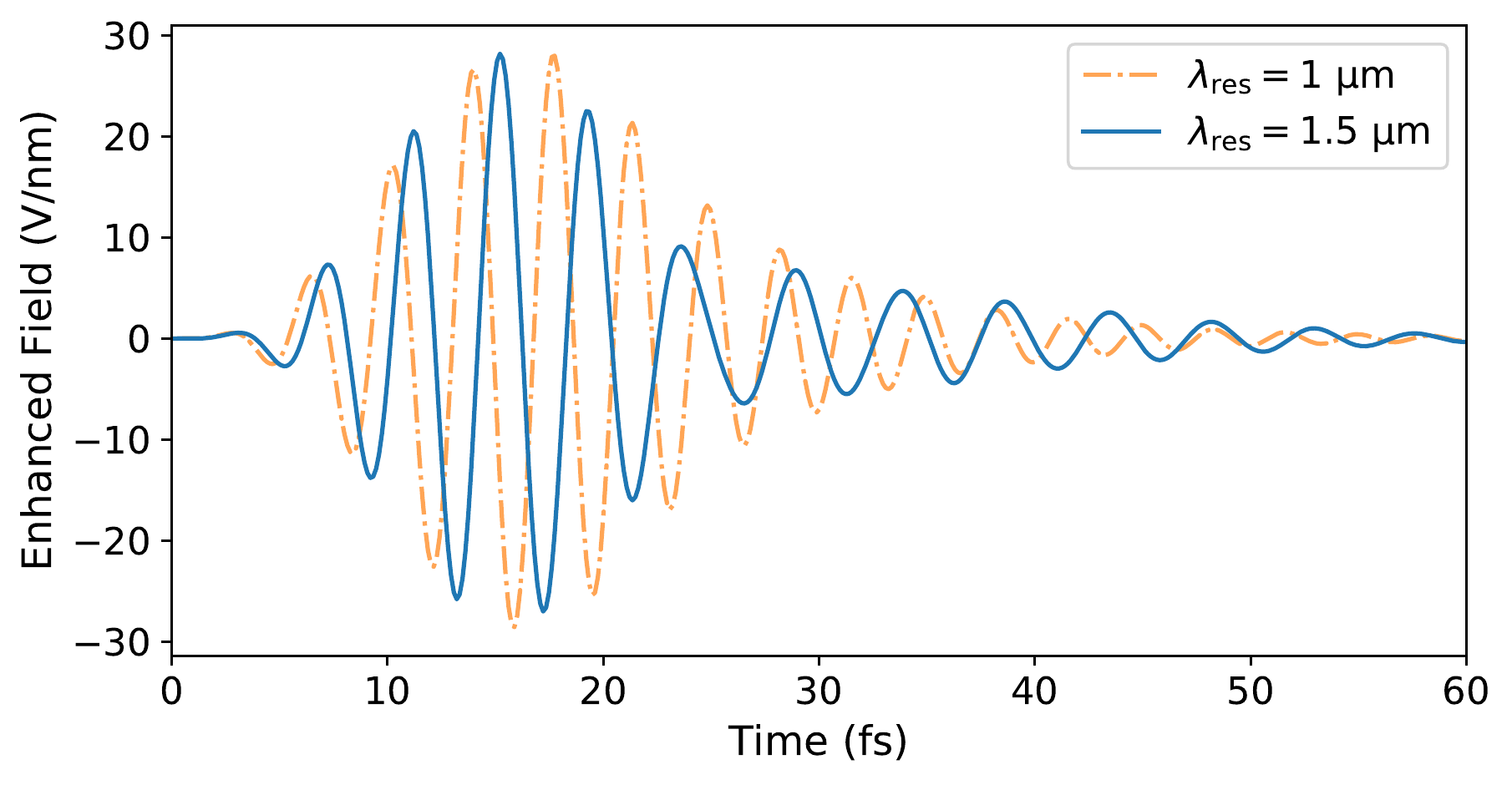}
		\caption{Comparison of enhanced electric field waveforms at the tip apex for $\lambda_\text{res} \approx \SI{1}{\micro\meter}$ (dashed orange) and $\lambda_\text{res} \approx \SI{1.5}{\micro\meter}$ (solid blue).  For both cases the input field was a $\cos^2$ waveform with a central wavelength of $\SI{1.177}{\micro\meter}$, a FWHM of \SI{10}{\femto\second}, and peak field strength of \SI{1}{\volt\per\nano\meter}.  While their peak field strengths and envelope durations are similar, the waveform excited at the apex of an antenna with $\lambda_\text{res} \approx \SI{1}{\micro\meter}$ lasts 2.83 optical cycles, while that excited at the apex of an antenna with $\lambda_\text{res} \approx \SI{1.5}{\micro\meter}$ lasts just 2.3 optical cycles.  This results in nearly an order of magnitude increase in $Q_1$ from the antenna with $\lambda_\text{res} \approx \SI{1.5}{\micro\meter}$ compared to the antenna with $\lambda_\text{res} \approx \SI{1}{\micro\meter}$.}
		\label{fig:lambda_res_tip_waveform_comparison}
\end{figure}

To see this more clearly, let's compare two nanoantennas: one with a resonance wavelength of $\lambda_\text{res} \approx \SI{1.5}{\micro\meter}$, which is near the long-wavelength spectral edge; and the other with a resonance wavelength of $\lambda_\text{res} \approx \SI{1}{\micro\meter}$, which is near the short-wavelength spectral edge.  Figure~\ref{fig:lambda_res_tip_waveform_comparison} shows the enhanced electric field waveform at the tip apex for both nanoantennas excited by the same \SI{10}{\femto\second} full-width at half-maximum (FWHM) $\cos^2$ source waveform having a source strength of \SI{1}{\volt\per\nano\meter}.  Given that two nanoantennas yield nearly identical peak field strengths (roughly \SI{30}{\volt\per\nano\meter}) and similar envelope durations, the average total photoemission $Q_0$  from the two antennas is roughly equal (see Figure~\ref{fig:average_total_emission_curved_teardrop}).   However, the nanoantenna transfer function effectively shifts the spectrum of the enhanced optical waveform at the nanoantenna apex relative to the incident optical waveform.  For $\lambda_\text{res} \approx \SI{1.5}{\micro\meter}$,  the central wavelength of the electric field waveform at the nanoantenna apex has been shifted up to \SI{1.22}{\micro\meter} with a FWHM pulse duration of roughly $2.3$ optical cycles, while for $\lambda_\text{res} \approx \SI{1}{\micro\meter}$, the central wavelength of the electric field waveform at the nanoantenna tip has been shifted down to \SI{1.127}{\micro\meter} with a FWHM pulse duration of roughly 2.83 optical cycles.  Due to the fewer number of optical cycles within the FWHM of the enhanced electric field waveform for $\lambda_\text{res} \approx \SI{1.5}{\micro\meter}$, the CEP-sensitive photocurrent $Q_1$ increases by roughly an order of magnitude compared to the case of $\lambda_\text{res} \approx 1 \si{\micro\meter}$ as shown in Figure~\ref{fig:cep_dependent_charge_curved_teardrop}.  Resonance wavelengths that fall further outside of the bandwidth of the optical source are not ideal as the field enhancement is significantly reduced, thereby leading to lower average total and net CEP-sensitive photoemission $Q_1$.  

\FloatBarrier

\begin{figure}[h]
		\centering
		\includegraphics[width=\textwidth]{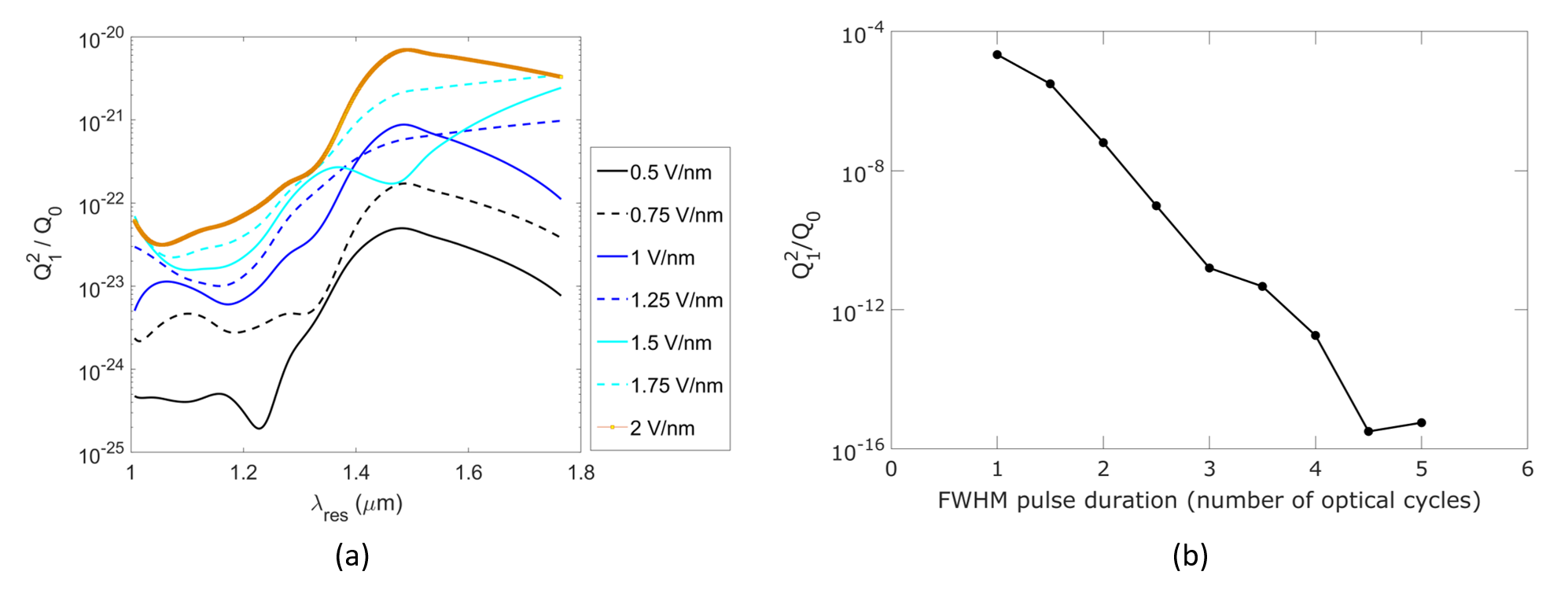}
		\caption{Investigation of SNR. a) Plot of $Q_1^2/Q_0$ which is proportional to the SNR as a function of the resonance wavelength for various different incident source strengths.  Improvements in SNR of up to $50$-$100\times$ are observed for antennas having resonances positionted toward the longer-wavlength regions of the source spectrum.  b) Plot of $Q_1^2/Q_0$ as a function of FWHM pulse duration in number of optical cycles from a single surface emission site.  Shorter cycle durations lead to dramatic enhancements in the SNR.  A local field strength of $\SI{14}{\volt\per\nano\meter}$ was assumed for this calculation, but similar results are observed for all field strengths tested.  Note that the units for this calculation are arbitrary as the constant pre-factors in the Fowler-Nordheim equation and surface integration were ignored.}
		\label{fig:SNR_curved_teardrop}
\end{figure}

Finally, we can use the our results to investigate the dependence of SNR of the CEP-sensitive signal defined by $Q_1^2/Q_0$.  In Figure~\ref{fig:SNR_curved_teardrop}a we plot the SNR as a function of resonance wavelength for several values of the peak incident field strength.  As expected, the SNR values are optimized for resonance wavelengths aligned with the long-wavelength edge of the source spectrum, with the SNR near $\lambda_\text{res} = \SI{1.5}{\micro\meter}$ being between $50\times$ to more than $100\times$ larger than for $\lambda_\text{res} = \SI{1.177}{\micro\meter}$ near the central wavelength of the source spectrum.  This is due to a combination of the increasing value of $Q_1$ and simultaneously decreasing values of $Q_0$ as the resonance wavelength approaches the long-wavelength edge of the source spectrum (see Figures~\ref{fig:average_total_emission_curved_teardrop} and \ref{fig:cep_dependent_charge_curved_teardrop}).  Following our discussion above, this is again a result of the compromise between the peak field strength and pulse duration in number of optical cycles.  To demonstrate this, we calculated $Q_1^2/Q_0$ assuming various FWHM pulse durations of enhanced surface waveforms each having an equivalent peak field strength of $\SI{14}{\volt\per\nano\meter}$ and central wavelength of $\SI{1.177}{\micro\meter}$. The results are plotted in Figure~\ref{fig:SNR_curved_teardrop}b as a function of waveform duration in number of optical cycles.  In general, the SNR depends strongly on the cycle duration of the optical waveform. As shown in the figure, a 2-cycle pulse leads to roughly 3.6 orders of magnitude higher SNR compared to a 3-cycle pulse. 

\FloatBarrier

%Figure~\ref{fig:simulation_cell2} shows the CEP sensitivity $Q_1/Q_0$ for various FWHM pulse durations of a transform limited pulse with a central wavelength of $\SI{1.177}{\micro\meter}$ illuminating a balanced differential emitter~\cite{rybka_sub-cycle_2016, ludwig2019sub, yang2019light}. The balanced differential emitter consists of two opposing planar gold surfaces, with the total emission current being the difference between the emission currents from the two surfaces. 
\section*{Conclusion}

We have investigated the impact of geometry, in particular sidewall curvature, and resonance wavelength on optical-field photoemission from resonant nanoantennas.  We were able to show that simple adjustments to the shape of the nanoantennas, in particular the addition of curved sidewalls leading to the nanoantenna apex, can result in very broadband increases in the peak electric field enhancement factor as a result of an enhanced lightning-rod effect.  Such geometric improvements in the electric field enhancement factor are favorable for ultrafast photoemission applications, such as CEP detection, as they preserve the ultrafast character of the enhancement which is critical to achieving high CEP-sensitivity.  %In contrast, an increase in quality factor might also be used to increase field enhancement at a particular wavelength, but compromises device bandwidth. 

We also studied impact of the nanoantenna's resonance wavelength, which was controlled by adjusting the nanoantenna size, on the CEP-dependent photoemission $Q_1$.  While the average total photoemission $Q_0$ was always greatest when the resonance wavelength $\lambda_\text{res}$ was equal to the central wavelength of the source waveform, the CEP-sensitive photoemission exhibited a nontrivial dependence on both $\lambda_\text{res}$ and the peak field strength of the source.  Despite this, we found that on average the per-device CEP-sensitive photoemission was greater when $\lambda_\text{res}$ was located near the long-wavelength edge of the source spectrum as opposed to aligned with the central wavelength of the incident pulse, as this condition struck the best balance between field enhancement and FWHM pulse duration in number of optical cycles of the enhanced tip waveform.  In general, we found that the FWHM duration of the enhanced waveform in number of optical cycles is a key metric for enhancing CEP sensitivity of nanoantenna optical-field emitters, further motivating investigations of few-cycle plasmonic excitation~\cite{csete_few-cycle_2020}.  Using a combination of curved sidewalls and optimized resonance wavelength tuning, we find that improvements of the CEP-sensitive photocurrent of $5$-$10\times$ and SNR of $50\times$ to greater than $100\times$ relative to a triangular antenna operated on-resonance.  

This work will inform the development of on-chip petahertz electronics for optical-field detection and signal processing.  Following on these findings, we anticipate that with further optimization such field-sensitive detectors can be improved to the extent that they will enable direct shot-to-shot CEP detection of low-energy few-cycle optical pulses in integratable, chip-scale platforms.  

%%%%%%%%%%%%%%%%%%%%%%% References %%%%%%%%%%%%%%%%%%%%%%%%%

%Add references with BibTeX or manually.
%\cite{Zhang:14,OSA,FORSTER2007,Dean2006,testthesis,Yelin:03,Masajada:13,codeexample}

%\cite{swanwick_nanostructured_2014,hobbs_high-yield_2014,hobbs_high-density_2014,musumeci_multiphoton_2010,rybka_sub-cycle_2016,rathje_review_2012,putnam_few-cycle_2019,putnam_optical-field-controlled_2017,paulus_measurement_2003,guler_refractory_2014,telle_carrier-envelope_1999,racz_measurement_2017,piglosiewicz_carrier-envelope_2014,paulus_absolute-phase_2001,paasch-colberg_solid-state_2014,kern_electrically_2015,jones_carrier-envelope_2000,herink_field-driven_2012,dombi_ultrafast_2013,chew_increasing_2019,keathley2019vanishing,fowler_electron_1928,forbes_description_2008}

\section*{Appendix A: Construction of Nanoantenna Geometries}
\label{sec:appendix_A}

\begin{figure}[h]
		\centering
		\includegraphics[width=\textwidth]{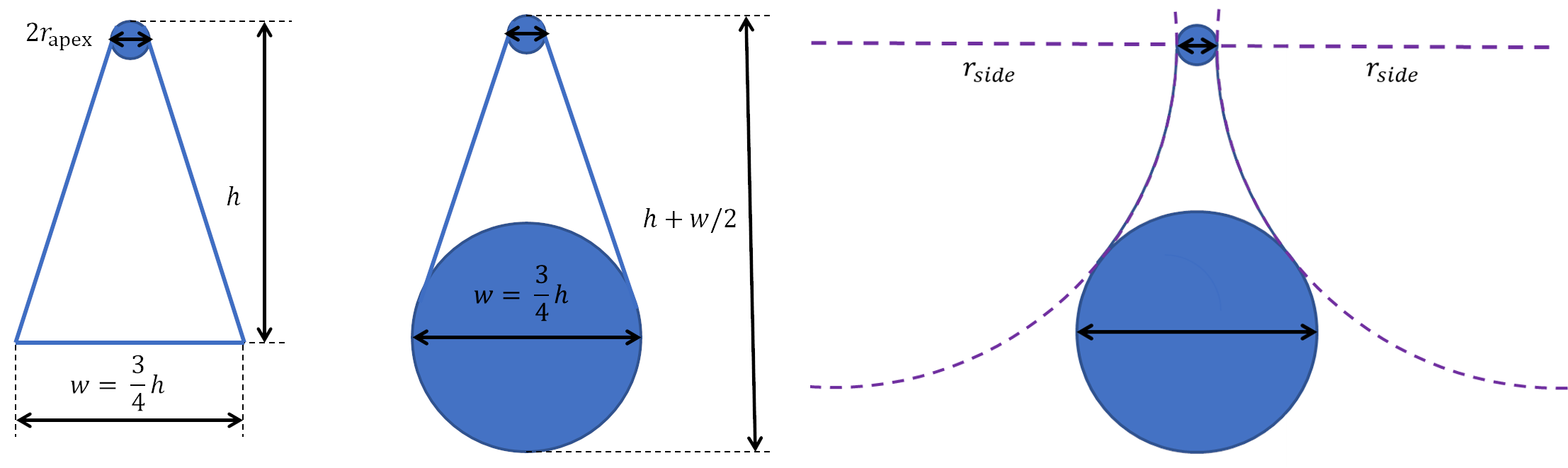}
		\caption{Construction of the different nanoantenna geometries.  From left to right: the triangular nanoantenna, the simple teardrop nanoantenna, and the curved teardrop nanoantenna.}
		\label{fig:nanotip_geometry}
\end{figure}

Here we provide further detail on the construction of each nanoantenna geometry in MEEP for the simulations presented in the manuscript.  A diagram of each geometry is provided in Figure~\ref{fig:nanotip_geometry}.  The triangular antenna was constructed first using an extruded triangle with the tip tip removed and replaced with a cylinder of radius $r_\text{apex}$.  The aspect ratio of this triangle was fixed to $h/w = 4/3$ for all simulations.  Starting from this geometry, the simple teardrop was then formed by simply adding a cylinder at the base of the triangle having a diameter equal to the base width $w$.  And finally, the curved teardrop was simply a modification to the curved teardrop whereby an empty cylinder was used to cut away from the sidewalls of the triangular tip leading to the apex.  The radius of the cylinder $r_\text{side}$ was determined such that it fell tangent to both the cylinder forming the apex and the base of the simple teardrop structure.  

\FloatBarrier

For resonance tuning, all dimensions of these structures except for $r_\text{apex}$ were simply proportionally increased or decreased in size in the $x$ and $y$ dimensions while maintaining the original aspect ratio of the core extruded triangle.  The tip radius of curvature $r_\text{apex}$ was left fixed at \SI{10}{\nano\meter} for all simulations as this was a reasonable value for typical fabrication resolution for such structures.  

\section*{Appendix B: CEP-Sensitivity}
\label{sec:appendix_B}

\begin{figure}[h]
		\centering
		\includegraphics[width=0.6\textwidth]{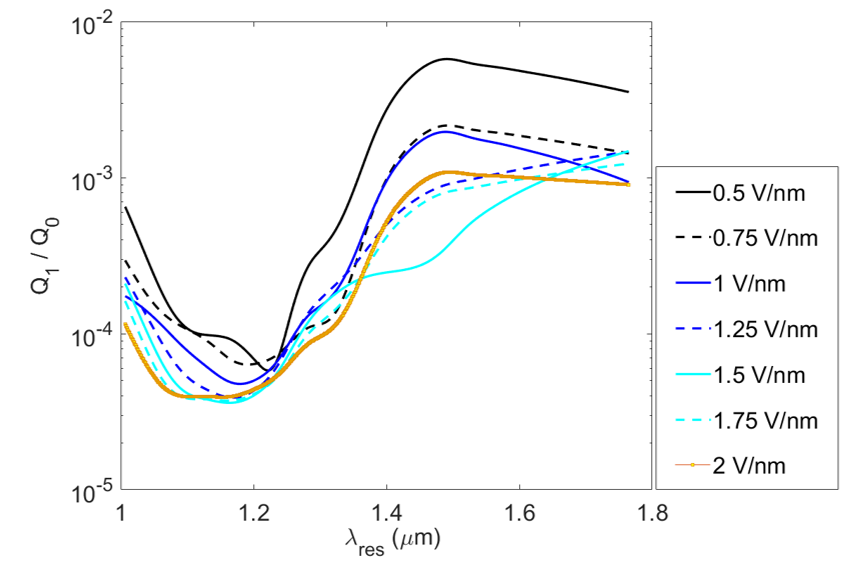}
		\caption{Study of the CEP-sensitivity $Q_1/Q_0$ of the curved teardrop structure as a function of the resonance wavelength.}
		\label{fig:cep_sensitivity}
\end{figure}

While in this report we focused on the SNR of the emitted CEP-sensitive photocurrent, several works in the literature report on the CEP-sensitivity $Q_1/Q_0$.  For the interested reader, we have included a plot of the resonance tuning study of the CEP-sensitivity for the curved teardrop structure (see Figure~\ref{fig:cep_sensitivity}) which mirrors Figure~\ref{fig:SNR_curved_teardrop} of the manuscript.  

\FloatBarrier

\section*{Acknowledgements}

This material is based upon work supported by the Air Force Office of Scientific Research under award numbers FA9550-19-1-0065 and FA9550-18-1-0436.  We thank Brenden Butters and Akshay Agarwal for their scientific discussion and edits to the manuscript.  

%%%%%%%%%% If using BibTeX:
\bibliographystyle{ieeetr}
\bibliography{biblio}

%%%%%%%%%% If preparing manually:
% \newcommand{\enquote}[1]{``#1''}

% \bibitem{Zhang:14}
% Y.~Zhang, S.~Qiao, L.~Sun, Q.~W. Shi, W.~Huang, L.~Li, and Z.~Yang,
%   \enquote{Photoinduced active terahertz metamaterials with nanostructured
%   vanadium dioxide film deposited by sol-gel method,}
%   {\protect\JournalTitle{Optics Express}} \textbf{22}, 11070--11078 (2014).

% {Optical Society}, \enquote{{OSA Publishing},}
%   \url{http://www.osapublishing.org}.

% \bibitem{FORSTER2007}
% P.~Forster, V.~Ramaswamy, P.~Artaxo, T.~Bernsten, R.~Betts, D.~Fahey,
%   J.~Haywood, J.~Lean, D.~Lowe, G.~Myhre, J.~Nganga, R.~Prinn, G.~Raga,
%   M.~Schulz, and R.~V. Dorland, \enquote{Changes in atmospheric consituents and
%   in radiative forcing,} in \enquote{Climate Change 2007: The Physical Science
%   Basis. Contribution of Working Group 1 to the Fourth assesment report of
%   Intergovernmental Panel on Climate Change,}  S.~Solomon, D.~Qin, M.~Manning,
%   Z.~Chen, M.~Marquis, K.~B. Averyt, M.~Tignor, and H.~L. Miler, eds.
%   (Cambridge University Press, 2007).

% \end{thebibliography}

\end{document}